**Exploratory Analysis of Multivariate Longitudinal Child Education Data**

Maria Vivien Visaya

Department of Pure and Applied Mathematics

University of Johannesburg[a]

v3isaya@gmail.com and mvvisaya@uj.ac.za

(corresponding author)

David Sherwell

School of Computer Science and Applied Mathematics,

University of the Witwatersrand, South Africa

David.Sherwell@wits.ac.za

Charles Kimpolo

School of Computer Science and Applied Mathematics,

University of the Witwatersrand, South Africa

Charles@aims.ac.za

Mark Collinson

Medical Research Council University of the Witwatersrand

Unit in Rural Public Health and Health Transitions Research

(The Agincourt Health and Population Unit),

PO Box 2 Acornhoek 1360[d]

Mark.Collinson@wits.ac.za

**Exploratory Analysis of Multivariate Longitudinal Child Education Data**

**ABSTRACT**

We analyse binary multivariate longitudinal data of a population of households from a rural district in South Africa. Using a 2-dimensional graphical representation of longitudinal data, each household's data is transformed into a time-evolving geometric orbit. Orbits communicate complete information of change in the data over time and provide insights into the dynamics of both a household's and the population's evolution. The outcome of interest is child educational default, where defaulting is defined as having failed more than three years of schooling. A visual analysis of the impact on educational default of three household factors, namely the presence of a biological mother, the age of the household head (minor- or adult- headed household) and the death of an adult, is presented. In both the non-defaulting and defaulting households, dynamics is mainly described by the temporary in- and out-migration of biological mothers. We find that the presence of mother is more likely in the non-defaulting households. Owing to insufficient events involving change in the age of the household head and adult deaths, we have no conclusion regarding their effect. Orbits offer easily interpreted information of clusters, patterns of change, and the density of state transitions of household orbits.

**Keywords:   data analysis; data visualization; multivariate data; educational progression**

# 1. INTRODUCTION

Longitudinal data is the repeated measurement of the same variables on the same subjects over time. Analysis of multivariate longitudinal data (MLD) is interesting as it offers the opportunity for cross-sectional analysis at any moment, or analysis of change in outcomes. Longitudinal surveys give detailed change so *frequency of answer change* is a property of interest (Singer and Willett, 2003). For analysis of MLD, one statistical approach is to construct transition models. For instance, binary data in *n* variables will require a $2^n$-state model and construction of transition probabilities between states (Gottschau, 1994; Ilk, 2008; Zeng and Coo, 2007). Regression techniques on the other hand involve parameter estimation for the explanatory variables (Bandyopadhyay et al., 2011).

Graphical techniques in longitudinal data are equally important (Tufte, 2001). Understanding the changes that have occurred in a longitudinal study over time, particularly those with huge amount of quantitative information are easier to interpret and understand in their visual format. For visual analysis of MLD, suggested methods are given in (Al-Aziz et al., 2010; Wang et al. 2014) but very few are available when data is binary. Here we use the method of orbits, a simple way of describing and analysing binary MLD first introduced in (Kimpolo, 2011) and developed in (Visaya and Sherwell, 2014). Orbits are useful in handling data sets with large number of variables and subjects. The advantage of using orbits over other methods such as the Markov chain model, generalized estimating equation model, motion charts, and heat map is discussed in (Visaya et al., 2015).

Our purpose is

$$\textit{"to study the effect of three household changes on child schooling"}. \tag{1}$$

Some literature that discuss household determinants affecting children's school progression are worth mentioning. In Matlab Bangladesh, male out-migration has a positive effect on child schooling but mother out-migration does not (Alam and Streatfield, 2009). Education of father is highly related to school progression in Cambodia (Keng, 2004) while in Sri Lanka, the presence of the mother is significant (Dissanayake et al., 2014). As for South Africa, the quality of education has been poor (Fleish, 2008). In our data, we find that about 80% of children will fail once in their schooling. In (Case and Ardington, 2006), it is found that the loss of a child's mother is a strong predictor of poor schooling outcomes. For rural Agincourt in South Africa, almost all children enroll at age seven but educational progress is often delayed with few post-secondary opportunities (Kahn et al., 2012).

With regards to our purpose, the effect on child's schooling is examined for three household variables, namely

1) mother's temporary out-migration,

2) age of household head (minor or adult), and

3) death of an adult in a household.

The following binary questions relate to educational progression and are associated respectively to the variables above:

1) $Q_0$ (BM): is the biological mother residing in the household?

2) $Q_1$ (HH): is the household head a minor?                          (2)

3) $Q_2$ (AD): has there been an adult death?

The data analysed here involves household variables that affect children of school-going age (7-16 years). Longitudinal data of each household is visualized by constructing its 2-dimensional orbit representation. In (Visaya and Sherwell 2014), the orbit method is analysed in mathematical detail and is briefly illustrated using the three questions in (2). However, detailed analysis and interpretation of orbits were not discussed. In addition, only complete data was considered. Here we give a detailed analysis of household orbits in the 2-dimensional state space and also accommodate missing data. Particular attention will be placed on the determination of the structure, characteristics, and dynamics of the population using patterns of orbits.

This paper is summarized as follows. In Section 2 we give a brief outline of the method of orbits. We work an example that will be followed throughout the section (from coded data to its step by step orbit construction). We give a description of the Agincourt household data in Section 3. The measure for both child default and mother's temporary migration are discussed as well. Orbit results for the defaulting and nondefaulting subpopulations are discussed in Section 4. We give concluding remarks in Section 5.

## 2. THE METHOD OF ORBITS

The method of orbits (Visaya and Sherwell 2014) considers a longitudinal data of binary questions from a population of subjects over time. For consistent interpretation of visuals, coding of answers is hypothesized as either favourable (coded 1) or unfavourable (coded 0) to a given purpose. For questions given in (2), a 'yes' to $Q_0$ is hypothesized favourable to educational progression, while a 'yes' to either $Q_1$ or $Q_2$ is unfavorable. By the 0/1 coding of answers, the set of $n$ responses at any moment can be represented by a single binary string of length $n$ and the data of the subject is described by a sequence of these strings. Table 1 illustrates an example of the coded answers of household $k$ to n=3 questions given in (2) at times t=0,1,…,7. Observe that answer to $Q_1$ is constant, while answer to $Q_2$ is frequently changing. These properties are not reflected in the line graph of the time series of $k$ shown in Figure 1.

Aside from using concatenated answers, the *order of variables* is also considered in constructing orbits. In particular, the order of variables is dynamically rearranged at each time, with the objective of extracting clusters of orbits in stable, least frequently changing variables. The use of variable order as an analytical tool is different from the analysis of question order and the psychological effect of question order in a survey (Schumann and Presser, 1981; Fox and Tracy, 1986) or from the design of the questionnaire (William, 1993; Burford et al., 2009).

| t | $Q_0$ (BM) | $Q_1$ (HH) | $Q_2$ (AD) |
|---|---|---|---|
| 0 | 1 | 1 | $1^{\#}$ |
| 1 | $1^{\#}$ | 1 | $0^{\#}$ |
| 2 | 0 | 1 | $1^{\#}$ |
| 3 | 0 | 1 | $0^{\#}$ |
| 4 | 0 | 1 | $1^{\#}$ |
| 5 | $0^{\#}$ | 1 | $0^{\#}$ |
| 6 | $1^{\#}$ | 1 | $1^{\#}$ |
| 7 | 0 | 1 | 0 |

**Table 1** Favourable(=1)/Unfavourable(=0) coding of data of *k* to questions in (2).

Answer with a number sign change value in the next time step.

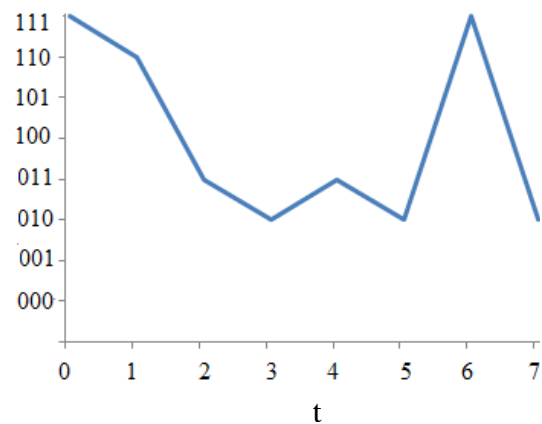

**Figure 1** Line graph of the time series of concatenated answers to fixed question order $Q_0, Q_1, Q_2$

$(= 012)$    in Table 1.

We briefly outline the orbit construction (Visaya and Sherwell, 2014). Let

I={0,1,2,…,n-1} : be an index set.

For each subject $k$ in the population, let

$$f_{i_j}^k \text{ be the frequency of change in answer value of}$$

question $i_j \in I$ over the observation period.

Suppose

$$0 < f_{i_0}^k < f_{i_1}^k < \cdots < f_{i_j}^k < f_{i_{(n-1)}}^k. \tag{3}$$

Given (3), the *initial question order of subject $k$* is

$$y_0^k = i_0 i_1 \ldots i_{n-1}, \tag{4}$$

where

a)  if $f_{i_j}^k = f_{i_{j+1}}^k$ and $f_{i_j}^{pop} < f_{i_{j+1}}^{pop}$ at the population level, then question order is chosen as $i_j i_{j+1}$.

b)  if $f_{i_j}^{pop} = f_{i_{j+1}}^{pop}$, then question order is chosen according to the trivial order 012…(n-1).

The *initial answer state of $k$ is*

$$x_0^k = x_0 x \ldots x_{n-1}$$

where $x_j$ is the corresponding answer to $i_j$ in $y_0^k$. The *initial state of $k$ is*

$$p_0^k = (x_0^k, y_0^k).$$

A simple algorithm for determining the next states $p_t^k = (x_t^k, y_t^k)_{t>0}$ of a subject $k$ is given by the following steps[1]:

<u>*Step 1*</u>: Determine initial state $p_0^k = (x_0^k, y_0^k)$ of $k$.

<u>*Step 2*</u>:  Identify the questions that change answer values at $t$=1.

a)  If there are none, then the next state $p_1^k = p_0^k$. Let

$$x_j^* = \begin{cases} 0 \text{ if } x_j = 1 \\ 1 \text{ if } x_j = 0 \end{cases}.$$

If answer of $Q_{i_j}$ changes at t=1, then swap $i_j$ and corresponding new answer $x_j^*$ to the rightmost position of the digit string. So for t=0 where

$$p_0^k = (x_0 x_1 \ldots x_j \ldots x_{n-1}, i_0 i_1 \ldots i_j \ldots i_{n-1}),$$

the new state at t=1 is $p_1^k = (x_1^k, y_1^k)$ where

$$x_1^k = x_0 x_1 \ldots x_{n-1} x_j^* \quad \text{and} \quad y_1^k = i_0 i_1 \ldots i_{n-1} i_j.$$

b) Suppose more than one answer changes at t=1. Then sequentially swap changing questions and corresponding new answers to the right, starting with the rightmost changing question. For instance, let both $Q_{i_j}$ and $Q_{i_{j'}}$ change answers at t=1, where j< j'. Denote new answers of $Q_{i_j}$ and $Q_{i_{j'}}$ by $x_j^*$ and $x_{j'}^*$ respectively. So from time t=0 where

$$p_0^k = (x_0 x_1 \ldots x_j \ldots x_{j'} \ldots x_{n-1}, i_0 i_1 \ldots i_j \ldots i_{j'} \ldots i_{n-1}),$$

the new state at t=1, is $p_1^k = (x_1^k, y_1^k)$ where

$$x_1^k = x_0 x_1 \ldots x_{n-1} x_j^* x_j^* \quad \text{and} \quad y_1^k = i_0 i_1 \ldots i_{n-1} i_{j'} i_j.$$

_Step 3:_   Update $p_0^k$ as $p_1^k$. Repeat process in Step 2 and iterate until $t$=T-1.

We note that Steps 1-3 is applied to *each* subject in the population. Longitudinal data of each subject is presented as a time-evolving *orbit* defined by the sequence of states

$$O(k) = \{p_0^k, p_1^k, \ldots, p_T^k\}.$$

The orbit method is simply a rearrangement of data at each step. It is an exact coding of data so all information is preserved. To follow the evolution of states, say from state $p$ to $p'$, we connect states $p$ and $p'$ with an edge. Transition from state $p$ to state $p'$ is denoted by $p \rightarrow p'$. Given the number of binary variables $n$, let $X_n$ be the set of all possible concatenated answers and let $Y_n$ be the set of all concatenated order of variables. The orbit of each subject in the longitudinal data is plotted in a 2-dimensional space

$$S_n = \ X_n \ \times \ Y_n$$

composed of $N = 2^n \times n!$ states. Figure 2 illustrates $S_n$ for n=3.

**Remark 1**   Let times $t \neq 0$, $t'$, and $r \geq 2$ be such that $t < t' < t+r$. Suppose that data of subject $k$ for $Q_i$ is missing at time interval *(t,t+r)*. The initial condition of $k$ is still defined using (4), where frequency $f_i$ is computed such that an answer changes if value of $Q_i$ at $t$ is different from value at $t+r$. For states $p_{t'}$ where $t' \in (t, t+r)$, we use last observation carried forward method (Xu, 2009) and assume that $p_{t'} := p_t$.

**Example.** We apply the orbit algorithm to the data of subject $k$ in Table 1.

<u>Step 1</u> : state $p_0^k = (x_0^k, y_0^k)$    Answer to Q1 is constant, while answers to Q0 and Q2 changed 3 times and 7 times respectively (changing answers denoted by number sign). The frequency relation (as in (3)) is

$$f_1^k < f_0^k < f_2^k.$$

The initial state of k is then composed of $y_0^k = 102$ and $x_0^k = 111$.

<u>Step 2:</u> state $p_1^k = (x_1^k, y_1^k)$     At t=1, only answer to Q2 changes (from 1 to 0). Using $p_0^k = (11\mathbf{1}, 10\mathbf{2})$, we swap both variable 2 and corresponding answer 1 to the right (they are both already on the right), and change answer 1 to 0. The next state is $p_1^k = (110, 102)$.

<u>Step 3:</u>   state $p_2^k = (x_2^k, y_2^k)$    At t=2, both answers to Q2 and Q0 change. Using Step 2c) and $p_1^k = (1\mathbf{10}, 1\mathbf{02})$, we    first swap variable 2 and corresponding answer 0 to the right (they are both already on the right), and change answer 0 to 1. Next, we swap to the right variable 0 with corresponding answer 1 to the right, and change answer 1 to 0. Hence we have  $p_2^k = (110, 120)$.

States of $k$ for t=1,2,…,7 are given in Table 2. The evolution of the orbit $O(k)$ in    $S_3$ is shown in Figure 2. States  $p_t^k$  and  $p_{t+1}^k$  are connected by an edge.

| $t$ | $x_t$ | $y_t$ | $t$ | $x_t$ | $y_t$ |
|---|---|---|---|---|---|
| 0 | $111^*$ | $102^*$ | 4 | $101^*$ | $102^*$ |
| 1 | $11^*0^*$ | $10^*2^*$ | 5 | $10^*0^*$ | $10^*2^*$ |
| 2 | $11^*0$ | $12^*0$ | 6 | $11^*1^*$ | $12^*0^*$ |
| 3 | $100^*$ | $102^*$ | 7 | $100$ | $102$ |

**Table 2**    States at each time t=0,1,…,7 associated to data of $k$ in Table 1. Answers (and corresponding variables) with asterisks change value at the next time step in Table 1.

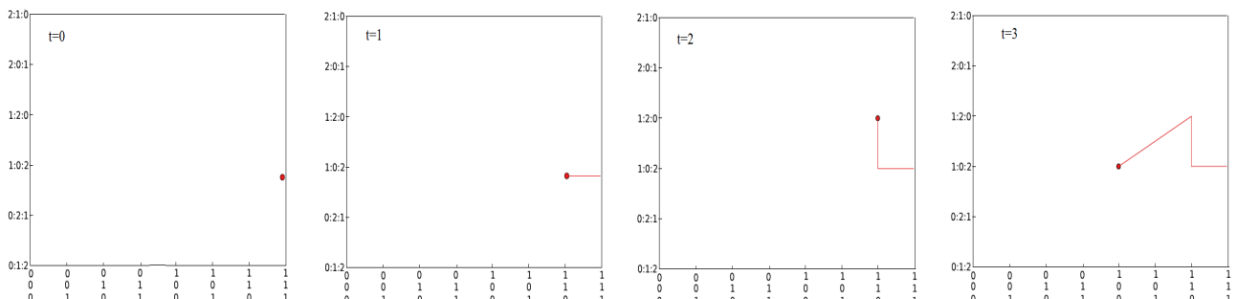

**Figure 2**   Evolution of orbit *O(k)* in $S_3$ associated to data in Table 1. The dot at each time t=0 to t=7 represents the state $p_t^k = (x_t^k, y_t^k)$. The y-axis is composed of *3!* possible orders of questions 0(BM), 1(HH), and 2(AD) while the x-axis is all $2^3$ binary strings of length 3.

To help in understanding the evolution of orbit in Figure 2, we note the information given by $p_0^k$=(111,102) in Table 2:

1. Variable 1 is at the leftmost position of $y_0^k$, with corresponding value 1 in $x_0^k$. This means that $Q_1$(HH) is the most stable variable, often favourable in value. We see from t=2 to t=7of Figure 2 that the orbit *O(k)* stays in the subset of $S_3$ where $Q_1$=1.

2. Variable 2 is at the rightmost position of $y_0^k$ which means that $Q_2$ is the least stable variable. Except for t=2 and t=6 in Figure 2, *O(k)* often visit states where the rightmost variable in $y_t^k$ is 2.

**Remark 2**

a) **Swapping to the far right** allows clusters associated to the least changeable variables to be seen in $S_n$. For K=3000 simulated orbits, n=4 variables, and time length T=10, Figure 3(a) illustrate the strategy of swapping to the right-end, while Figure 3(b) swaps to the left-end. In Figure 3(a), clusters in the left half (resp. right half) of $S_4$ are unfavourable (resp. favourable) in the leading variable. In Figure 3(b), clustering of orbits is lost.

**b) Initial order** $p_0^k$ based on (4) enables faster clustering in $S_n$. For K=3000 simulated orbits, n=13 variables, and T=10, Figure 4 illustrates states of orbits with initial condition using (4) while Figure 5 illustrates evolution of states starting with the same initial condition. The state of orbits for each time t= 0,1,..., 10 is denoted by dots.

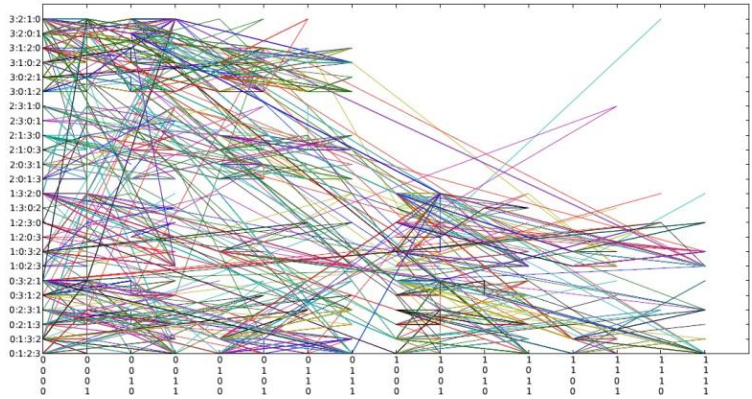

(a)

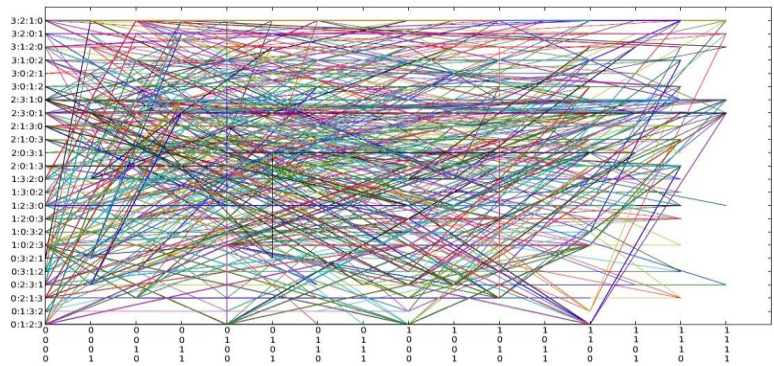

(b)

**Figure 3** Simulated orbits in $S_4$ using strategy of (a) swapping to the right-end and (b) swapping to the left-end. Clusters are lost in (b).

Observe from Figure 4 that clustering of orbits at t=0 remain clustered until t=10. In Figure 5, all orbits are given the same initial order using the population frequency relation so $y_0 = 2{:}1{:}0{:}3{:}6{:}9{:}7{:}8{:}5{:}4{:}10{:}12{:}11$ (i.e. at the population level, $f_2 < f_1 < \ldots < f_{11}$). Transitions (edges) from previous times are retained. The location of orbits in Figure 4 at t=0 is similar to the location of orbits of Figure 5 at t=10. This informs us that the choice of initial significance using (4) places orbits in their expected cluster.

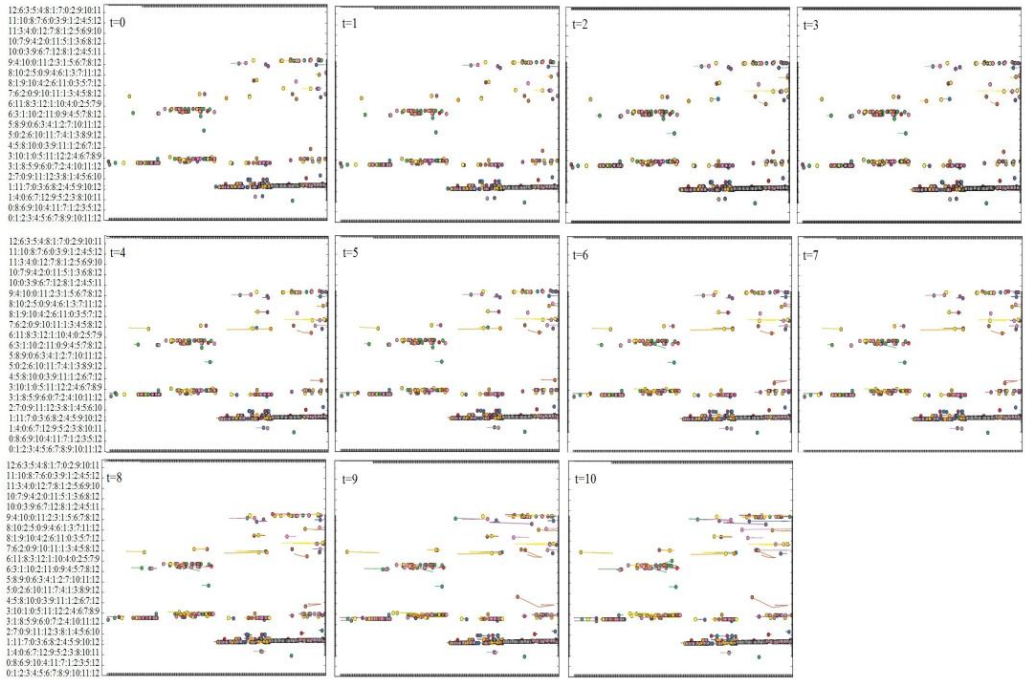

**Figure 4**   States of 3000 simulated orbits in $S_{13}$ for times t=0 to t=10 (denoted by dots) with initial significance for each orbit using (4).

## 3.   AGINCOURT HOUSEHOLD DATA

The Agincourt Health and Demographic Surveillance Site (HDSS) constitutes a sub-district of Bushbuckridge district, Mpumalanga Province and covers 402 km$^2$. It is located in the remote, north-east of South Africa, close to the eastern border of Mozambique. It is rural, with poor infrastructure and services. Almost a third of the study population consists of Mozambican immigrants. Temporary (labour) migration maintained strongly high rates for men (60% in men in the age group 35-54 years) and growing proportions of adult women (from around 5% of women in the age group 15-34 in 1997 to 19% in 2001). Close to half (43%) of the women who became temporary migrants in 1999 or 2000 had at least one child. The proportion of female-headed households increased significantly from 29% in 1992 to 33% in 2000. A more detailed history and evolution of the Agincourt study population has been described in (Tollman, 1999; Tollman et al. ,1999, 1995; Kahn et al. , 2007). The Agincourt HDSS maintains a relational database that is a longitudinal representation of annual population data captured and upgraded through Microsoft Structured Query Language (SQL) Server 2005. The baseline of the HDSS was established in 1992. The data collected is based on a repeated census taken on the 31[st] of

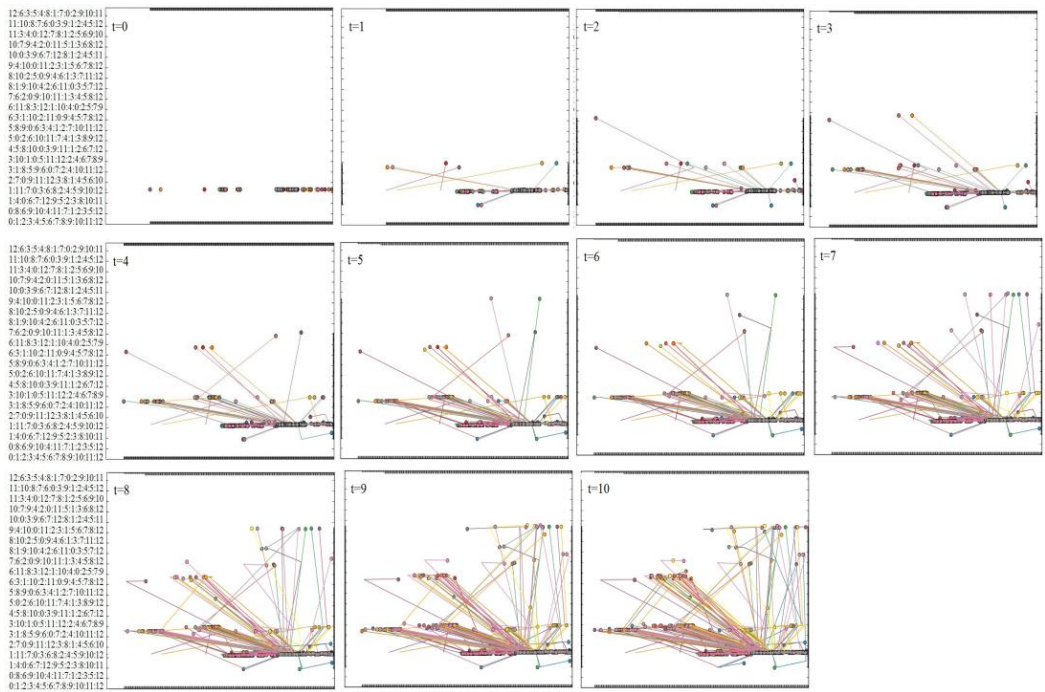

**Figure 5**     States of 3000 simulated orbits in $S_{13}$ for times t=0 to t=10 (denoted by dots). All orbits are started at the same initial significance y=2:1:0:3:6:9:7:8:5:4:10:12:11.

December of each when data is available. Figure 6 illustrates the percentage (of the total Agincourt household population) of households that have been observed at each year from 1992 to 2007. More than 60% of the population has data available every year from 1998.

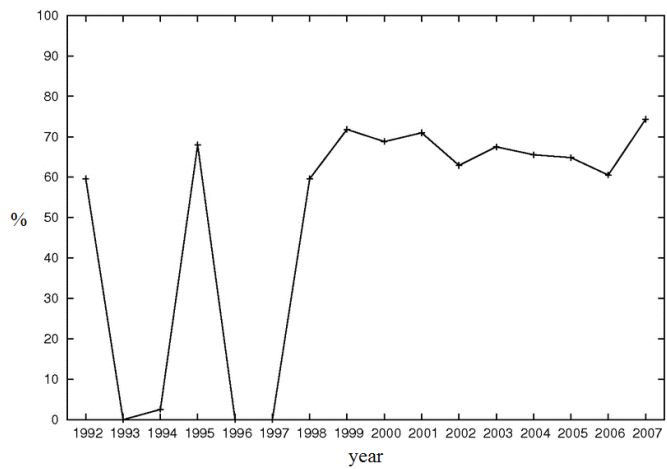

**Figure 6**   Percentage of households (out of the total Agincourt household population) with available data each year from 1992 to 2007.

The duration of data analysed here is from 1998-2007 and includes K=2,699 households with children of school-going age (7-16 years old). Household identifiers are anonymized or de-identified by the data managers to ensure confidentiality. Recall the purpose given in (1) and the 0/1 coding of three binary questions given in (2), namely

$$Q_0(BM): yes=1, \quad Q_1(HH): yes=0, \text{ and } \quad Q_2(AD): yes=0.$$

Regarding $Q_0(BM)$, the absence/presence of a mother in the household is measured by the number of months that she has resided in the household (Collinson, 2010). We record mother as

**absent** :    if mother has resided in the household for less than 6 months of the survey year.

**present** :    if mother has resided in the household for at least 6 months of the survey year.

If there are children in the household with different biological mothers, with at least one mother having been absent for the survey year, then answer to $Q_0$ is 'no'.

With regards to data on school progression, education status data had only been available for the five years 1992, 1997, 2002, 2005 and 2008. The available information for each child in the household are ID, age, and the total number of completed years of education.

For each child, let

$\mathbf{f}$ : denote the number of education years that

the child failed during his school life, and

$\mathbf{a}$ : denote the age of the child.

We associate the number of completed education years to the age of the child by the function

$$\mathbf{y} : \mathbf{a} \rightarrow \mathbf{y(a)}.$$

For the period 1998−2007, the average observation time is 7 years so we cannot measure whether or not a child of age $\mathbf{a} < (7 + \mathbf{f})$ has failed any particular grade $\mathbf{f}$ times. We assume that a child at age $\mathbf{a} < (7 + \mathbf{f})$, regardless of the number of education years $\mathbf{y(a)}$, is *not* a defaulting child. Now for other ages $\mathbf{a} \geq (7 + \mathbf{f})$ we have the following:

1) if $\mathbf{y}(7 + \mathbf{f} + \mathbf{r}) > (\mathbf{f} - 1) + \mathbf{r}$, then child is non-defaulting

2) if $\mathbf{y}(7 + \mathbf{f} + \mathbf{r}) \leq (\mathbf{f} - 1) + \mathbf{r}$, then child is defaulting

where $\mathbf{r}$ varies with respect to the age range. For the Agincourt data, the age range is from 7 to 16 years. Here, we assume that $\mathbf{f} \geq 2$, so

$$\mathbf{f} - 2 \leq \mathbf{r} \leq 7.$$

The distribution of the fraction of defaulting Agincourt households with respect to $\mathbf{f}$ is illustrated in Figure 7.[2]

On average, the Agincourt population is significantly defaulting. This is a severe criticism of the quality of education offered in Agincourt. We find that the majority of the population defaults for **f**=2 and **f**=3 but for **f**=4, there is roughly a balance in the two populations, i.e. 54.17% defaulting and 45.82% of non-defaulting. We classify households with at least one child that have failed more than three education years (i.e. **f**=4) during their school life as *defaulting households*. Otherwise they are *non-defaulting*.

---

[2] The code for educational default can be viewed in http://issc.uj.ac.za/~vivien/ (choose the "code for default or nondefault" txt file from the "codes and data for Agincourt Child Education" zip file).

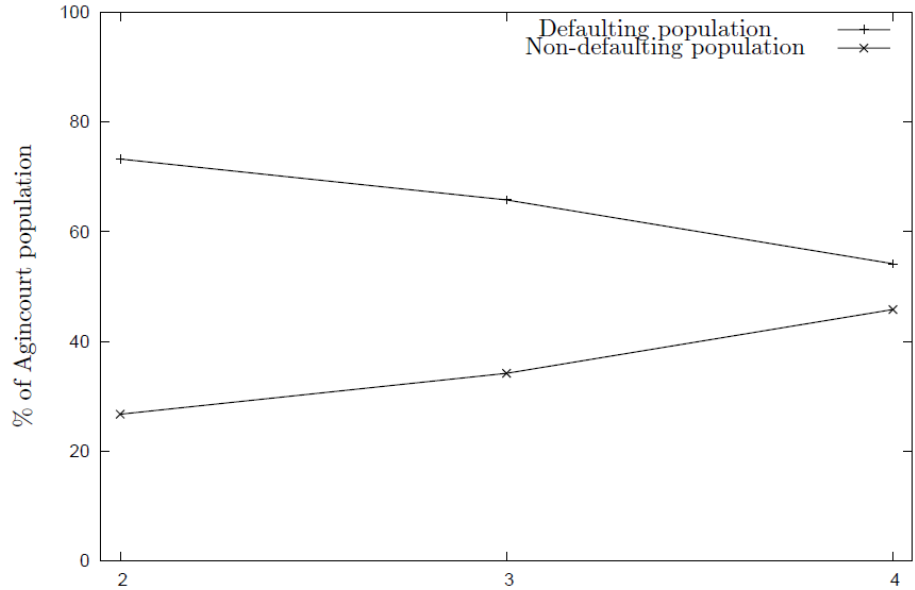

**Figure 7**   Distribution of the Agincourt defaulting and non-defaulting households with respect to the number of child education failure years **f**.

## 4. ORBIT RESULTS

The population frequencies per variable are

$$f_1^{pop}=0.16\% <\ f_2^{pop}=13.12\% < f_0^{pop}=86.72\%.$$

The variable household head (HH) associated to $Q_1$ is very (favourably) stable which means that there are very few households in Agincourt that are headed by minors. On the other hand, biological mother (BM) changes most often. It is interesting to note that at the population level, adult death is less stable than HH. These are immediate crude observations. Figures 8(a) and 8(b) illustrate orbits of non-defaulting and defaulting households in a subspace of $S_3$, respectively. We immediately can visualize and identify clusters. For graphic representation of longitudinal data, it is important to include time as an additional variable. Figures 8(c) and 8(d) show the household orbits over time. Automatic clustering of orbits serves as a visual cue to guide us to region(s) of interest (e.g. the region with many transitions).

Denote by

$d_{ij}$ :    the accumulated number of   transitions from state i to state j

which we refer to as the *density* from i to j.

$d_{ij}^{def}$: the density from i to j in the defaulting population.

$d_{ij}^{nondef}$: the density from i to j in the non-defaulting population.

Figures 9(a) and 9(b) show $d_{ij}$ involving states from the set

$$H=\{23, 24, 29,30,31,32\}$$

$$=\{(x,y) : x=1^{**}, y=1^{**}\}$$

for the non-defaulting and defaulting sub-populations respectively. We note that most of the transitions (~97.67%) happen in H, where every household is headed by an adult (i.e. $Q_1(HH)=1$). Stable adult household head is a characteristic of the stable (darker) clusters in Figures 8(c) and 8(d).

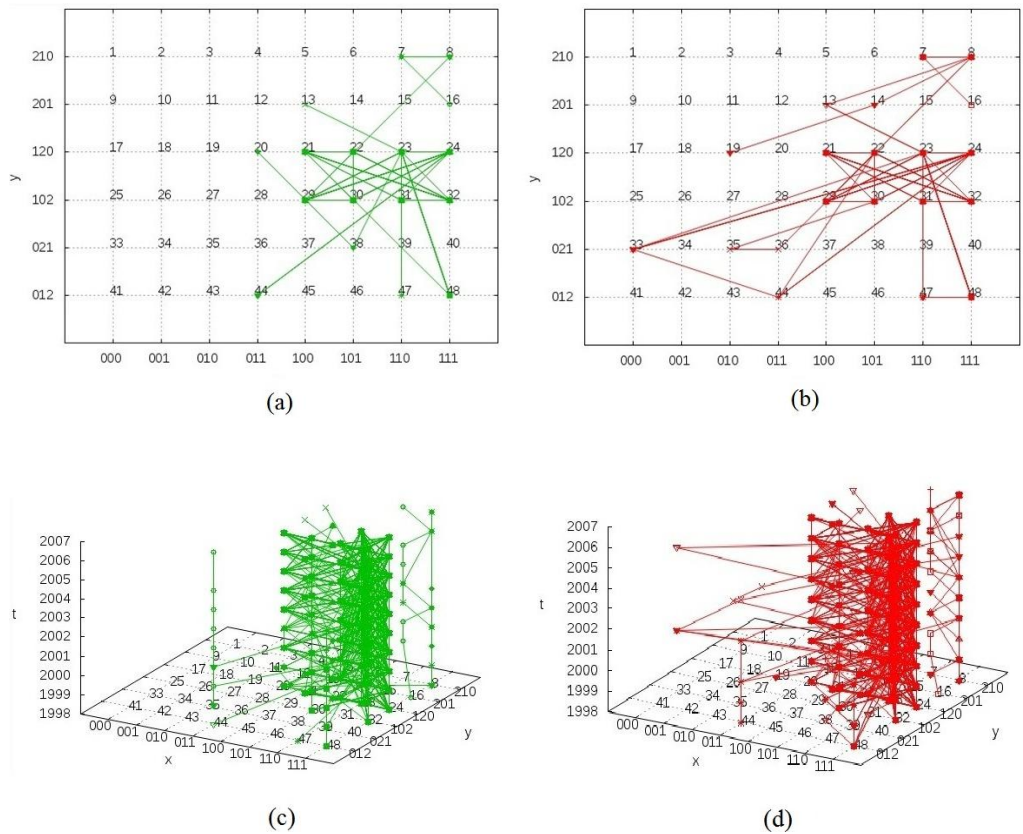

**Figure 8**    Orbits of (a) non-defaulting and (b) defaulting households in S*3*. Orbits of (c) non-defaulting and (d) defaulting households over time.

In demography, we note the possibility to simplify dynamics in S*3* by ignoring small densities. Figure 9 shows that for both defaulting and non-defaulting subpopulations, $d_{ij}$ is dominant in the subset

$$L=\{23=(110,120),\ 24=(111,120)\}$$

where both $Q_1$(HH) and $Q_2$ (AD) are favourably constant (i.e. households headed by an adult and no adult death). The entire analysis then comes down to transitions involving states 23 and 24.

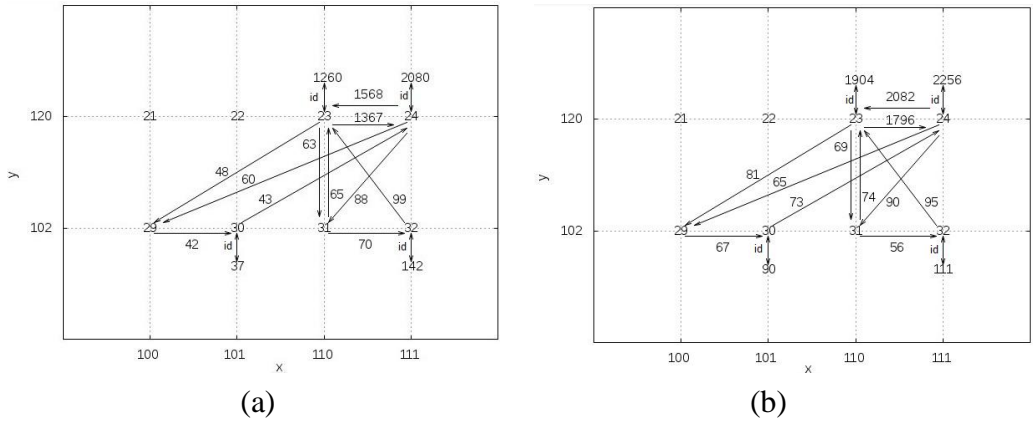

(a)  (b)

**Figure 9**   Accumulated dominant transitions $d_{ij}$ in H={23, 24, 29,30,31,32} for the (a) non-defaulting and (b) defaulting Agincourt households orbits. In this subset of S$_3$, we have the stable variable Q$_1$(HH)=1. Density with "id" means density of transition to and from the same state (i.e. $d_{ij}$ where i=j).

**Remark 3**

a)   The flow of orbits in L is described by temporary migration of biological mothers. State transitions 23→ 24 are characterised by in-migration of mother (from absent to present), while 24→ 23 is characterized by out-migration (from present to absent). Orbits that idle in 23 have mothers away for the whole survey period, while those that idle in 24 have mothers living in the household over the survey period.

b)   Although we only have one significant demographic variable, our conclusions are a subtle improvement in Markov analysis. Markov analysis selects a single y-value (arbitrary fixed variable order). Say we choose fixed question order as y=120 in Figure 9. Then the information given in state 29 is the same as that given in sate 21 (which we denote by 29=21), 31=22, 30=23, and 32=24. Transitions are then collapsed, e.g. density in 23→24 and 30→ 24 in $S_3$ are combined in this fixed question order. This effect of "unfolding" states in $S_n$ brought by including order of variables show significant transitions that are lost in using a fixed question order.

Figure 10 gives the number of households at each year in each of the 8 states i=21, 22, 23, 24, 29, 30, 31, 32. The dominant numbers are associated to states 23 and 24, with graphs being symmetrical. In particular, an increase in absent mothers (increasing graph of 23) results in the decrease in present mothers (decreasing graph for 24). These two states are unaffected by the other states.

Denote the density from *i* to *j* in the population by

$$d_{ij}^{P} = d_{ij}^{def} + d_{ij}^{nondef}.$$

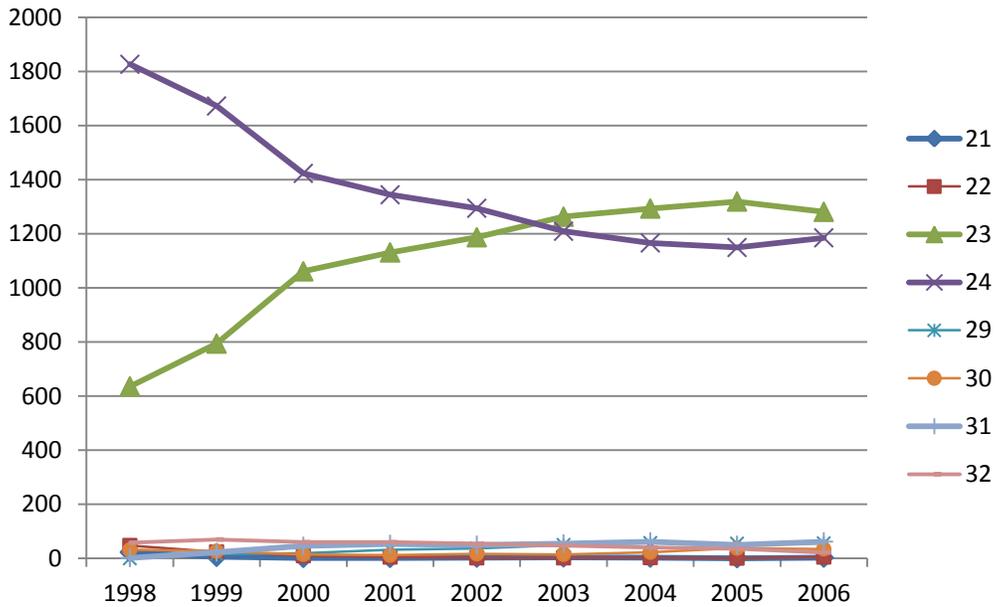

**Figure 10**    Number of household orbits in state *i*    (*i*=21, 22, 23, 24, 29, 30, 31, 32)    for each year of the survey period. Note the increase in mothers who temporarily out-migrate (increasing graph for 23).

Table 3 gives the number and proportion of accumulated transitions from $i \rightarrow j$ that occurs in the non-defaulting and defaulting households (percentage computed as $d_{ij}^{nondef}/d_{ij}^{P}$ and $d_{ij}^{def}/d_{ij}^{P}$, respectively). Fix state pair $i, j$ and let states $i' \neq i, j' \neq j$. Define

**a** $= d_{ij}^{nondef}$ : number of transitions $i \rightarrow j$ in non-defaulting households,

**b** $= \sum_{i',j'} d_{i'j'}^{nondef}$ : number of transitions $i' \rightarrow j'$ in non defaulting households,

**c** $= d_{ij}^{def}$ : number of transitions $i \rightarrow j$ in defaulting households, and

**d** $= \sum_{i',j'} d_{i'j'}^{def}$ : number of transitions $i' \rightarrow j'$ in defaulting households.

The **odds ratio** (OR) of transition from state i to state j is given by

$$\textbf{OR} = \frac{\frac{a}{c}}{\frac{b}{d}} = \frac{ad}{cb}$$

Where **OR>1** means that transition is more likely in the non-defaulting households, **OR<1** means that transition is more likely in the defaulting households, and **OR=1** gives no conclusion.

Although Figures 8 and 9 illustrate many similar transitions and clusters in both sub-populations, we have the following observations from Table 3:

1. For the non-defaulting households, the dominant transition is 24→ 24 (~48% of all transitions 24→ 24 in the population). This suggests that households in this very favourable state (HD=1, AD=1, BM=1) are more likely to stay in this condition. This is supported by the odds ratio of 1.26, i.e. mothers are more likely to stay at home in the non-defaulting households.

2. For the defaulting households, the transition 23 → 23 is dominant (~60% of all transitions 23→ 23 in the population). This means that once mothers become absent from residence, they take long to return to their households. This confirms with the odds ratio of 0.79, which means that mothers are more likely to be away from home in the defaulting households.

3. In-migration (absent to present) dynamics dominates in non-defaulting households while out-migration (present to absent) dynamics dominates in defaulting households. This is seen in the dominant transitions 24→24 and 23→24 (transitions ending in mother present) in the non-defaulting households, while it is 24→23 and 23→ 23 (transitions ending in mother absent) for the defaulting households.

**Table 3** Accumulated transitions $d_{ij}$ in L={23,24} for the non-defaulting and defaulting subpopulations.

| Transition | Non-defaulting | Defaulting | OR |
|---|---|---|---|
| 24 → 24 | 2080 | 2256 | **1.26** |
| (mother present to present) | *47.97%* | 52.03% | |
| 24→ 23 | 1568 | 2082 | 1.00 |
| (mother present to absent) | 42.96% | *57.04%* | |
| 23 → 23 | 1260 | 1904 | **0.79** |
| (mother absent to absent) | 39.83% | *60.17%* | |
| 23→ 24 | 1367 | 1796 | 1.05 |
| (mother absent to present) | *43.22%* | 56.78% | |

## 5. DISCUSSIONS

We analysed household survey data in three variables by using a graphical display of binary multivariate longitudinal data called orbits. The visualization is easily understood. Because information of change is carried by orbits, household dynamics is clear in $S_3$. The entire analysis comes down to transitions involving L={23,24}. In L, the phenomena of adult death and minor head household are not at all relevant to educational progression (nor are they coupled to mother temporary migration). Only mother migration is found to be defensible for our data. Our results contribute to the understanding of the role of household structures, particularly the role of the mother in providing a supportive home environment in which children can progress in terms of education. Educational progression will presumably be improved by policy that facilitates stay-at-home mothers. The state of education at Agincourt is obviously seriously defective and what we have presented is a useful accessible result for policy makers.

Concerning cause and effect, we cannot say that out-migration of mother causally precedes educational default. If transitions 23→24 and 24→24 were more dominant in the nondefaulting orbits, and 24→23, and 23→23 dominant in the defaulting orbits, then we might reasonably claim a causal chain. In any case, cause of mother migration is an obvious next step in understanding educational default. Of course many other phenomena may affect child's educational progression. For future work, it will be good to include other household factors, for instance unobserved factors such as household resources that is possible cause for absent mothers and/or educational default.

As for the method of orbits, it is a simple, useful, and informative tool for graphic representation and exploration of binary multivariate longitudinal data that involves many variables and subjects. Visual results present initial insights to help and complement statistical data analysis (e.g. Gelman, 2004; Garcia et al, 2004). For instance, density information from Figure 9 can be used in Markov transition models over the states in H (or L).

## ACKNOWLEDGEMENT

The data used in this study was supplied by the MRC/Wits Rural Public Health and Health Transitions Research Unit (Agincourt). The Agincourt HDSS is funded by the Medical Research Council and University of the Witwatersrand, South Africa, Wellcome Trust, UK (grant no. 058893/Z/99/A, 069683/Z/02/Z, 085477/Z/08/Z), and National Institute on Aging of the NIH (grants 1R24AG032112-01 and 5R24AG032112-03).

## REFERENCES

Alam, N., and Streatfield, P. K. (2009),    Parents' migration and children's education in Matlab, Bangladesh. Collinson, Adazu, White and Findly (eds). *The dynamics of migration, health and livelihoods. INDEPTH Network perspectives.* Surrey: Ashgate.

Al-Aziz, J., Christou, N., and Dinov, I. (2010), SOCR Motion Charts: An Efficient, Open-Source, Interactive and Dynamic Applet for Visualizing Longitudinal Multivariate Data. *Journal of Statistics Education*, 18(3).

Bandyopadhyay S., Ganguli B., and Chatterjee, A. (2011), A review of    multivariate longitudinal data analysis. *Statistical Methods in Medical Research*, 20 (4) : 299-330.

Burford, B. et al. (2009), Asking the right questions and getting meaningful responses: 12 tips on developing and administering a questionnaire survey for healthcare professionals. *Medical Teacher*, 31(3):207–211.

Case, A. and Ardington, C. (2006), The Impact of Parental Death on School Outcomes: Longitudinal Evidence From South Africa. *Demography,* 43 (3) : 401–420.

Collinson, M,. (2010), Striving Against Adversity: the Dynamics of Migration, Health and Poverty in Rural South Africa. *Global Health Action*, 3(2010).

Dissanayake, P., Chandrasekara, N., and Jayasundara, D. (2014), The Impact of Mother's

Migration for Work Abroad on Children's Education in Sri Lanka. *European International Journal of Science and Technology*, 3(1):110-120.

Fleisch, B. (2008), *Primary Education in Crisis: Why South African Schoolchildren Underachieve in Reading and Mathematics*. Cape Town: Juta & Co.

Garcia, R.E. et al (2004), Visual Analysis of Data from Empirical Studies, International Workshop on Visual Languages and Computing, Sept 2004.

Gelman, A. (2004), Exploratory Data Analysis for Complex Models, Journal of Computational and Graphical Statistics, 13(4): 755-779.

Gottschau, A. (1994), Markov Chain Models for Multivariate Binary Panel Data. *Scandinavian Journal of Statistics* 21(1):57-71.

Fox, J.A., and Tracy, P. (1986), *Randomized Response: A Method for Sensitive Surveys*. Sage Publications, Inc.

Ilk,O. (2008), *Multivariate Longitudinal Data Analysis: Models for Binary Response and Exploratory Tools for Binary and Continuous Response*. VDM Verlag.

Kahn,K. et al (2007), Research Into Health, Population and Social Transitions in Rural South Africa: Data and Methods of the Agincourt Health and Demographic Surveillance System. *Scandinavian Journal of Public Health*, 35(69): pp. 8-20.

Kahn,K. et al (2012), Profile: Agincourt Health and Socio-demographic Surveillance System. *International Journal of Epidemiology*, 41:988–1001.

Keng, C. (2004), Household Determinants of Schooling Progression Among Rural Children in Cambodia. *International Education Journal*, 5(4): 552-561.

Kimpolo, C. L. , Deterministic dynamics in questionnares in the social sciences, PhD Thesis (2011) University of the Witwatersrand, http://wiredspace.wits.ac.za/handle/10539/9248

Schuman, H. and Pressser, S. (1981), Questions and Answers in Attitude Surveys. *Quantitative Studies in Social Relations*. Academic Press, Inc.

Singer, J., and Willett, J. (2003), *Applied Longitudinal Data Analysis: Modeling Change and Event Occurrence*. Oxford University Press.

Tollman, S.M. et al (1999), The Agincourt Demographic and Health Study-site Description, Baseline Findings and Implications, *S Afr Med J.,* 89(8): 858-64.

Tollman, S.M. (1999), The Agincourt field site: Evolution and current status. *S Afr Med J*, 89:853–8.

Tollman, S.M., Herbst,K., and Garenne, M. (1995), The Agincourt Demographic and Health Study, Phase 1. *Johannesburg: Health Systems Development Unit, Department of Community Health, University of the Witwatersrand*.

Tufte, E.R., (2001) *The Visual Display of Quantitative Information* (2nd ed.), Cheshire, CT: Graphics Press.

Visaya, M.V., and Sherwell, D. (2014), Dynamics from multivariable longitudinal data. *Journal of Nonlinear Dynamics*, doi:10.1155/2014/901838

Visaya M.V., Sherwell, D., Sartorius, B., and Cromieres, F. (2015), Analysis of    Binary Multivariate Longitudinal    Data via 2-Dimensional Orbits : An Application to the Agincourt Health and Socio-Demographic Surveillance System in South Africa, PLoS One, 2015 Apr 28; 10(4):e012381 doi:10.1371/journal.pone.0123812.

Wang, F. et al (2014), What's In a Name? Data Linkage, Demography and Visual Analytics, M. Pohl and J. Roberts (eds),    EUROGRAPHICS 2014.

William, F. (1993), *Constructing Questions for Interviews and Questionnaires : Theory and Practice in Social Research*. Cambridge University Press.

Xu, H. (2007), LOCF Method and Application in Clinical Data Analysis, Conference Proceedings, NESUG 2009, Burlington, VT.

Zeng, L. and Cook, R. (2007), Transition models of multivariate longitudinal binary data. *Journal of the American Statistical Association,* 102(477).